# SENSOR SELECTION SCHEME IN TEMPERATURE WIRELESS SENSOR NETWORK

Mohammad Alwadi[1] and Girija Chetty[2]

[1]Department of Information Sciences and Engineering, The University of Canberra, Canberra, Australia
u3019769@uni.canberra.edu.au
[2]Faculty of ESTM, The University of Canberra, Canberra, Australia
girija.chetty@canberra.edu.au

## ABSTRACT

*In this paper, we propose a novel energy efficient environment monitoring scheme for wireless sensor networks, based on data mining formulation. The proposed adapting routing scheme for sensors for achieving energy efficiency from temperature wireless sensor network data set. The experimental validation of the proposed approach using publicly available Intel Berkeley lab Wireless Sensor Network dataset shows that it is possible to achieve energy efficient environment monitoring for wireless sensor networks, with a trade-off between accuracy and life time extension factor of sensors, using the proposed approach.*

## KEYWORDS

*Wireless sensor Networks, Physical environment Monitoring, machine learning, data mining, feature selection, adaptive routing.*

## 1. INTRODUCTION

Wireless and wired sensor networks (WSNs/SNs) has become a focus of intensive research today, especially for monitoring and characterizing of large physical environments, and for tracking environmental or physical conditions such as temperature, pressure, wind and humidity. A wireless or a wired sensor network (WSN/SN) consists of a number of sensor nodes (few tens to thousands) storing, processing and re-laying the sensed data, often to a base station for further computation [1, 2]. Sensor networks can be used in many applications, such as wildlife monitoring [3], military target tracking and surveillance [4], hazardous environment exploration [5], and natural disaster relief [6]. Many of these applications are expected to run unattended for months or years. Sensor nodes are however constrained by limited resources, particularly in terms of energy. Since communication is one order of magnitude more energy consuming than processing, the design of data collection schemes that limit the amount of transmitted data is therefore recognized as a central issue for wireless sensor networks. An efficient way to address this challenge is to approximate, by means of mathematical models, the evolution of the measurements taken by sensors over space and/or time. Indeed, whenever a mathematical model may be used in place of the true measurements, significant gains in communications may be obtained by only transmitting the parameters of the model instead of the set of real measurements. Since in most cases there is little or no a priori information about the variations taken by sensor measurements, the models must be identified in an automated manner. This calls for the use of machine learning and data mining techniques, which allow modelling the variations of future measurements on the basis of past measurements.

In this paper, we introduce a novel data mining based formulation for energy efficient WSN (Wireless Sensor Network) monitoring. The proposed approach involves an adaptive routing scheme to be used for energy efficiency and is based on selecting most significant sensors for the accurate modelling of the WSN environment. The experimental validation of the proposed scheme for publicly available Intel Berkeley lab Wireless Sensor Network dataset shows it is indeed possible to achieve energy efficiency without degradation in accurate characterization and understanding of WSN environment. The proposed machine learning and data mining formulation for achieving energy efficiency, provides better implementation mechanism in terms of trade-off between accuracy and energy efficiency, due to an optimal combination of feature selection and classifier techniques used in machine learning chain. By approaching the complexity of WSN/SN with a data mining formulation, where each sensor node is equivalent to an attribute or a feature of a data set, and all the sensor nodes together forming the WSN/SN set up - equivalent to a multiple features or attributes of the data set, it is possible to use powerful feature selection, dimensionality reduction and learning classifier algorithms from machine learning/data mining field, and come up with an energy efficient environment monitoring system [7]. In other words, by employing a good feature selection algorithm along with a good classification algorithm, for example, it is possible to obtain an energy efficient solution with acceptable characterization or classification accuracy (where the WSN/SN set up is emulated with a data set acquired from the physical environment). Here, minimizing the number of sensors for energy efficiency is very similar to minimizing the number of features with an optimal feature selection scheme for the data mining problem. Further, the number of sensors chosen by the feature selection scheme leads to a routing scheme for collecting the data from sensors, transmitting them until it reaches the base station node. As accuracy of data mining schemes rely on amount of previous data available for predicting the future state of the environment, it is possible to obtain an adaptive routing scheme, through the life of WSN, as more and more historical data becomes available, allowing trade-off between energy efficiency and prediction accuracy. We have shown that it is possible to do this, with an experimental validation of our proposed scheme with a publicly available WSN dataset acquired from real physical environment, the Intel Berkeley Lab [8]. Rest of the paper is organized as follows. Next Section describes the details of the dataset used, and Section 3 de-scribes the proposed machine learning - data mining approach. The details of experimental results obtained are presented in Section 4, and the paper concludes in Section 5, with conclusions and plan for further research.

## 2. DATA SET DESCRIPTION

The publicly available data set used for experimental validation consists of temperature and humidity measurements, which come from a deployment of 54 sensors in the Intel research laboratory at Berkeley [8]. The deployment took place between February 28th and April 5th, 2004. A picture of the deployment is provided in Figure 1 below, where sensor nodes are identified by numbers ranging from 1 to 54.

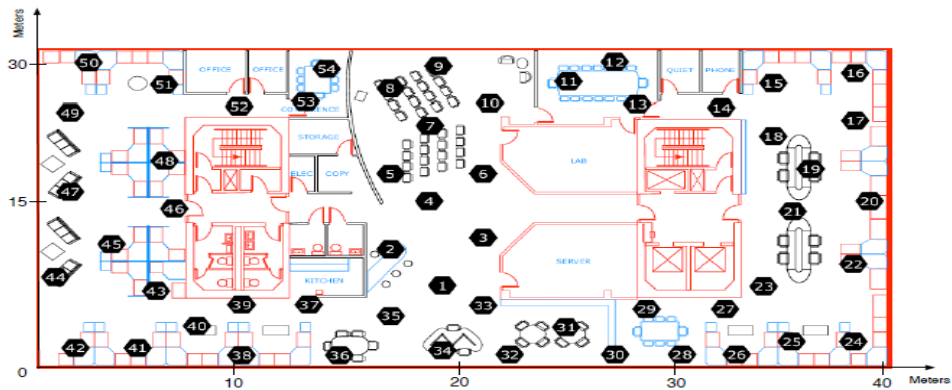

Figure 1. Intel Berkeley Wireless Sensor Network Data set: location of 54 sensors in an area of 1200m2

Many sensor readings from WSN test bed were missing, due to this being a simple prototype. We selected from this data set few subsets of measurements. The readings were originally sampled every thirty-one seconds. A pre-processing stage where data was partitioned was applied to the data set. After pre-processing, we prepared several subsets of data. The approach we used for this WSN test bed, involves, an assumption that data collection and transmission is done by some of the sensors (source nodes), that purely sense the environment and transmit their measurement to collector nodes (sink nodes/base station), and based on the relative distance between source nodes and sink nodes, the route or the path taken for sensor data to be transmitted from one node to other is predetermined at sink/base station node. Those nodes who actively participate in sensing the environment, and transmit the data, consume the power and those who do not participate in this activity do not consume any power. This is how the WSN can be made energy efficient; by involving optimum number of sensors to participate in environment sensing and transmission task, and leaving non-participating sensors in sleep mode (no energy consumption). This can however, impact on the accuracy of sensing the environment, if number of sensors participating in routing scheme is not properly chosen. To ensure a trade-off between accuracy and energy efficiency is achieved, it is essential that a dynamic or adaptive routing scheme is used, where, the machine learning/data mining technique can use larger training data from previous/historical data sets to predict the future environment accurately, and continuously adapt the routing scheme for nodes based on a threshold error measure for prediction accuracy and energy efficiency. Next Section describes the proposed machine learning data mining scheme used for sensor selection and routing in this approach.

## 3. SENSOR SELECTION AND ROUTING APPROACH

The sensor selection and routing approach is based on a feature selection technique that selects the attributes (sensors), by evaluating the worth of a subset of attributes by considering the individual predictive ability of each feature/sensor along with the degree of redundancy between them. Subsets of features that are highly correlated with the class while having low inter correlation are preferred [9, 10]. Further, this feature/sensor selection algorithm identifies locally predictive attributes, and iteratively adds the attributes with the highest correlation with the class as long as there is not already an attribute in the subset that has a higher correlation with the attribute in question. Once the appropriate group of sensors are selected, the prediction of sensor output at sink node or base station is done by linear regression algorithm, using the Akaike criterion [10, 11], which involves stepping through the attributes, removing the one with smallest standardised coefficient until no improvement is observed in the estimate of the error given by Akaike information metric. Figure 2 and the table below shows how the sensor selection evolves as the training data (historic data) used for predicting the sink sensor output is increased, and ensures the prediction accuracy/error is maintained at a particular threshold value.

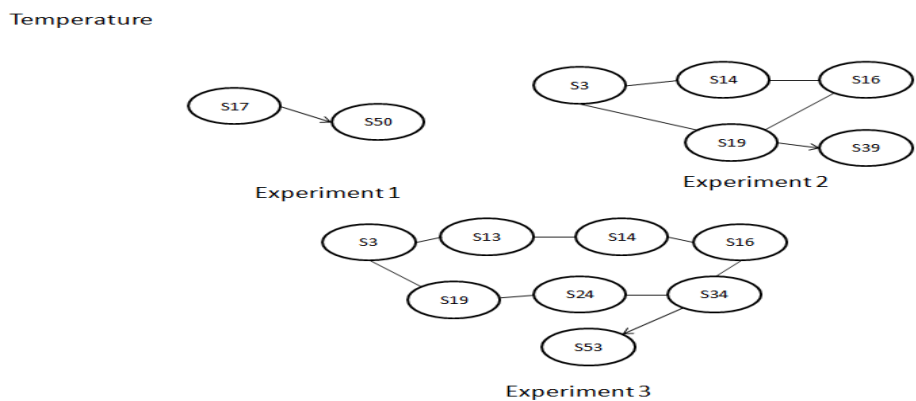

Figure 2. Sensor selection map for Temperature Experiment 1,2 and 3

## 4. EXPERIMENTAL RESULTS AND DISCUSSION

Different sets of experiments were performed to examine the relative performance of sensor selection and routing approach proposed here. We used k-fold stratified cross validation technique for performing temperature data set experiments, with k=2, 5 and 10, based on the training data available (using larger folds for larger training data). Further, to estimate the relative energy efficiency achieved, we performed experiments with all sensors (without feature selection/sensor selection) algorithm, and with sensors selected by feature selection algorithm. As mentioned before, the feature selection algorithm allows selection of an optimal number of features or sensor nodes needed to characterize or to classify the environment (which in turn leads to an energy efficient scheme). Further, time taken to build the model is also an important parameter, particularly for adaptive sensor routine scheme to be used for real time environment monitoring.

Table 1. Results for temperature Three experiments scenarios

| Exp number | Number of Sensors | Number of Training samples | Features/sensors selected | |
|---|---|---|---|---|
| 1 | 54 | 35 | 17,50 | |
| 2 | 53 | 2700 | 3,14,16,19,39 | |
| 3 | 53 | 5400 | 3,13,14,16,19,24,34,53 | |
| Exp number | Time (No F selection) | Time (F Selection) | RMSE No F selection | RMSE with F selection |
| 1 | 0.02 sec | 0.01 sec | 20.26% | 0.04% |
| 2 | 0.43 sec | 0.02 sec | 5.02% | 2.23% |
| 3 | 0.57 sec | 0.03 sec | 3.93% | 2.93% |

For the first set of experiments, we used first 54 sensors and a small set of training samples (35 temperature measurements). As can be seen from the sensor locations shown in Figure 2, sensor 17 and 50 are the sink node (emulating base station node), and sensors 1 to 54 participate in measuring and transmitting the environment around them to the sink node, where the machine learning prediction task is to estimate the measurement at sink node (sensor 50). The RMS error (root mean squared error) at the sink node (node 50) provides a measure of prediction For all source sensor nodes (1-54) in WSN participating in measuring the temperature in the environment and sending it to sink node, the RMS error is 20.26%, and with sensor selection scheme used with only 2 sensors participating in routing scheme, the RMS error is 0.04%. As can be seen in Table 1, with a moderate degradation in accuracy (20.26% to 0.04%), energy efficiency achieved is of the order of 54 (54/2). We used a new measure for energy efficiency, the life time extension factor (LTEF), which can be defined as:

$$\text{LTEF} = \frac{\text{Total number of sensors}}{\text{Sensors participating in the routing scheme}}$$

With 2 sensors out of 54 sensor nodes in active mode, the LTEF achieved is around 27 times, and 52 sensor nodes are in sleep mode. The trade off is a slight reduction in accuracy. This could be due to less training data used. We used only 35 temperature samples for prediction scheme. With more data samples used in the prediction scheme, performance could be better. To test this hypothesis, we performed next set of experiments.

For second set of experiments, we used 2700 training samples collected on different days. As can be seen in Table 1, with larger training data size, we found that the participating sensors in the routing scheme are different, as the proposed feature selection algorithm chooses different set of sensors (3,14,16,19,39). We used 53 sensors for this set of experiments, as one of the sensors (sensor 5 did not have more than 35 measurements). With all 53 sensors in the routing scheme, the RMS errors is 0.96%, and with 5 sensor nodes (3,14,16,19,39), the error is 5.02%. This was not a significant improvement in prediction accuracy (from 5.02% to 2.23%), with life time extension of 10.6 (53/5). As is evident here, by using larger training data (2700 temperature measurements), it was possible to achieve an improvement in prediction accuracy and energy efficiency as well.

To examine the influence of increasing training data size, we performed third set of experiments with 5400 samples. The performance achieved for this set of experiments is shown in Table 1. Here the adaptive routing scheme based on proposed feature selection technique selects 8 sensors (3,13,14,16,19,24,34,53). For this set of experiments, the RMS error varies from 3.93% for all sensors participating in the scheme to 2.93 % with LTEF of 6.6 (53/8). Though there is no degradation in prediction accuracy, there is not much improvement in energy efficiency, with doubling of training data size for the building the model this could be due to overtraining that has happened, with the network losing its generalization ability. So by increasing training data size, it may not be just possible to achieve performance improvement, for pre-diction accuracy (RMS error) and energy efficiency (LTEF), and a trade off may be needed. An optimal combination of training data size, and number of sensors actively participating in routing scheme can result in energy efficient WSN, without compromising the prediction accuracy.

Further, another important parameter is model building time, as for adaptive sensor routing scheme to be implemented in real time WSN environment, routing scheme has to dynamically compute the sensors that are in active mode and in sleep mode. Out of 3 experimental scenarios considered here, as can be seen from Table 1, the model building time from 0.02 seconds to 0.01 seconds for experiment 1, from 0.43 seconds to 0.02 seconds for experiment 2, and from 0.57 seconds to 0.03 seconds for experiment 3. So, the proposed adaptive routing scheme for sensor selection provides an added benefit of reduced model building times, suitable for real time deployment. Figure below shows the time taken to build the model.

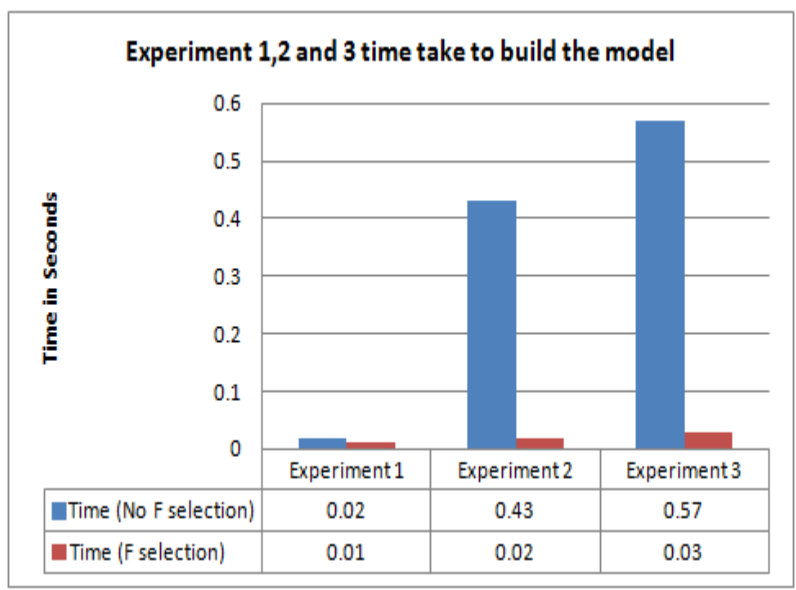

Figure 3. Time taken to build the model (Temperature Exp1,2 and 3)

# 5. CONCLUSIONS

Energy sources are very limited in sensor networks, in particular wireless sensor net-works. For monitoring large physical environments using SNs and WSNs, it is important that appropriate intelligent monitoring protocols and adaptive routing schemes are used to achieve energy efficiency and increase in lifetime of sensor nodes, with-out compromising the accuracy of characterizing the WSN environment. In this paper, we proposed an adaptive routing scheme for sensor nodes in WSN, based on machine learning data mining formulation with a feature selection algorithm that selects few most significant sensors to be active at a time, and adapts them continuously as time evolves. The experimental validation for a real world publicly available temperature WSN dataset, proves our hypothesis, and allows energy efficiency to be achieved without compromising the prediction accuracy, with an added benefit in terms of reduced model building times. Further work involves, developing new algorithms for sensor selection and environment characterization with WSNs and their experimental validation with other similar datasets, that can lead to better energy efficiency. Also, our further re-search involves extending this work with adapting these classifiers for big data stream data mining schemes, for real time dynamic monitoring of complex and large physical environments in an energy efficient manner.

## REFERENCES

[1]  Ping, S., Delay measurement time synchronization for wireless sensor networks. Intel Research Berkeley Lab, 2003.

[2]  Hall, M., et al., The WEKA data mining software: an update. ACM SIGKDD Explorations Newsletter, 2009. 11(1): p. 10-18.

[3]  Csirik, J., P. Bertholet, and H. Bunke. Pattern recognition in wireless sensor networks in presence of sensor failures. 2011.

[4]  Nakamura, E.F. and A.A.F. Loureiro, Information fusion in wireless sensor networks, in Proceedings of the 2008 ACM SIGMOD international conference on Management of data2008, ACM: Vancouver, Canada. p. 1365-1372.

[5]  Bashyal, S. and G.K. Venayagamoorthy. Collaborative routing algorithm for wireless sensor network longevity. 2007. IEEE.

[6]  Richter, R., Distributed Pattern Recognition in Wireless Sensor Networks.

[7]  Alwadi, M. and G. Chetty, Energy Efficient Data Mining Scheme for Big Data Biodiversity Environment. 2014.

[8]  Bodik, P., et al., Intel lab data. Online dataset, 2004.

[9]  Hall, M.A., Correlation-based feature selection for machine learning, 1999, The University of Waikato.

[10]  Akaike, H., Information theory and an extension of the maximum likelihood principle, in Selected Papers   of Hirotugu Akaike1998, Springer. p. 199-213.

[11]  Ashraf, M., et al., A New Approach for Constructing Missing Features Values. International Journal of Intelligent Information Processing, 2012. 3(1).

**Authors**

Mohammad Alwadi

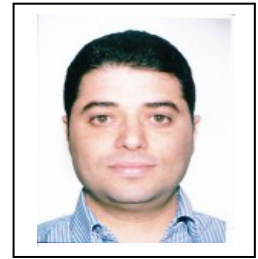

Mohammad Alwadi has bachelors degree in Computer Science from Jordan, Masters degree from the University of Canberra, Australia and Currently a PhD candidate in the Faculty of Information Science and Engineering at The University of Canberra, Australia. Mohammad research interests are in the area of computer networking, Data Mining and wireless sensor networks This author has 8 publications and working towards the submission of his PhD Thesis at the University of Canberra. Australia. He has more than 6 years of experience in the research field of Data mining , sensors networks and wireless sensor networks.

Dr.Girija Chetty

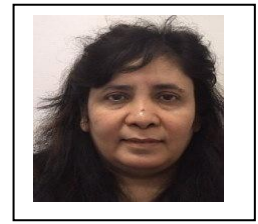

Girija has a Bachelors and Masters degree in Electrical Engineering and Computer Science from India, and PhD in Information Sciences and Engineering from Australia. She has more than 25 years of experience in Industry, Research and Teaching from Universities and Research and Development Organisations from India and Australia, and has held several leadership positions including Head of Software Engineering and Computer Science, and Course Director for Master of Computing (Mainframe) Course. Currently, she is an Associate Professor, and Head of the Multimodal Systems and Information Fusion Group in University of Canberra, Australia, and leads a research group with several PhD students, Post Docs, research assistants and regular International and National visiting researchers. She is a Senior Member of IEEE, USA, and senior member of Australian Computer Society, and her research interests are in the area of multimodal systems, computer vision, pattern recognition, data mining, and medical image computing. She has published extensively with more than 120 fully refereed publications in several invited book chapters, edited books, high quality conference and journals, and she is in the editorial boards, technical review committees and regular reviewer for several IEEE, Elsevier and IET journals in the area related to her research interests. She is highly interested in seeking wide and interdisciplinary collaborations, research scholars and visitors in her research group.